\documentclass[acmsmall]{acmart}

\AtBeginDocument{%
  \providecommand\BibTeX{{%
    \normalfont B\kern-0.5em{\scshape i\kern-0.25em b}\kern-0.8em\TeX}}}

\usepackage{subfigure}
\usepackage{multirow}
\usepackage{multicol}
\usepackage{algorithm2e}

\begin{document}

\title{Reversible Data Hiding in Encrypted Images Based on Bit-plane Compression of Prediction Error}

\author{Youqing Wu}
\authornote{Both authors contributed equally to this research}
\email{wuyq@hfnu.edu.cn}
\orcid{}
\affiliation{%
  \institution{Anhui Provincial Key Laboratory of Multimodal Cognitive Computation, School of Computer Science and Technology, Anhui University}
  \city{Hefei}
  \state{Anhui}
  \country{China}
  \postcode{230601}
}
\affiliation{%
  \institution{School of Computer Science and Technology, Hefei Normal University}
  \city{Hefei}
  \state{Anhui}
  \country{China}
  \postcode{230601}
}

\author{Wenjing Ma}
\authornotemark[1]
\email{e19201024@stu.ahu.edu.cn}
\author{Yinyin Peng}
\email{528456061@qq.com}
\author{Ruiling Zhang}
\email{zhangrling99@gmail.com}
\author{Zhaoxia Yin}
\authornote{Corresponding author}
\email{yinzhaoxia@ahu.edu.cn}
\affiliation{%
  \institution{Anhui Provincial Key Laboratory of Multimodal Cognitive Computation, School of Computer Science and Technology, Anhui University}
  \city{Hefei}
  \state{Anhui}
  \country{China}
  \postcode{230601}
}

\renewcommand{\shortauthors}{Wu and Ma, et al.}

\begin{abstract}
  As a technology that can prevent the information from being disclosed, the reversible data hiding in encrypted images (RDHEI) acts as an important role in privacy protection and information security. To make use of the image redundancy and further improve the embedding performance, a high-capacity RDHEI method based on bit-plane compression of prediction error is proposed in this paper. Firstly, the whole prediction error is calculated and divided into blocks of the same size. Then, the content owner rearranges the bit-plane of prediction error by block and compresses the bitstream with the joint encoding algorithm to reserve room. Finally, the image is encrypted and the information can be embedded into the reserved room. On the receiver side, the information extraction and the image recovery are performed separably. Experimental results show that the proposed method brings higher embedding capacity than state-of-the-art RDHEI works.
\end{abstract}



\maketitle

\section{Introduction}
In the past several decades, with the increasing demand for information security, data hiding technology \cite{Singh2020data} has been widely studied. As an essential part of data hiding technology, reversible data hiding (RDH) \cite{shi2016reversible} ensures that the embedded information is extracted completely and the original image is recovered correctly. At present, many traditional plaintext RDH methods have been proposed, which are mainly divided into three categories:
lossless compression \cite{zhang2013recursive}, difference expansion \cite{tian2003reversible}, and histogram shifting \cite{ni2006reversible}. These methods focus more on the imperceptibility and security of the embedded information, so the original image changes inconspicuously while embedding information. In other words, the information of the original image will be directly exposed in public. This can be harmful to some scenarios where privacy is essential. For example, the disclosure of image information on medical diagnostics may cause distress to patients.

Encryption algorithm converts plaintext information into unreadable ciphertext information. When the key is unknown, it is difficult for attackers to obtain the original content of the image. Therefore, the combination of the encryption algorithm and RDH can solve information leakage in traditional plaintext RDH method, thereby meeting the demands of privacy protection and information security. Specifically, the content owner encrypts the image and transmits it to the information hider. Then, the information hider embeds additional information in the encrypted image. Finally, the legitimate receiver extracts the information or recovers the image. This type of RDH technology combined with the encryption algorithm is called reversible data hiding in encrypted images (RDHEI) \cite{puteaux2021survey}. The RDHEI method can not only protect the information about the original image but also realize the transmission of additional information, which has a good application prospect in medical, military and judicial fields. Meanwhile, RDHEI also plays an important privacy protection role in cloud storage.

With the spread of cloud storage technology, plenty of RDHEI methods \cite{puech2008reversible,zhang2014reversibility,huang2016new,zhang2016lossless,xu2016separable} have been developed in recent years. Existing RDHEI methods can be mainly divided into two categories: vacating room after encryption (VRAE) \cite{zhang2011reversible,zhang2011separable,Qian2015Reversible,pun2020reversible,Wang2021High}, reserving room before encryption (RRBE) \cite{ma2013reversible,puech2008reversible,Mohammadi2020High,Guan2020efficient,Puteaux2021Recursive,Chen2021Multi}. 

In the first category, the VRAE method utilizes the redundancy of the encrypted image to vacate room for embedding information. Zhang proposed a block-splitting method of the encrypted image in  \cite{zhang2011reversible}. In this work, three least significant bits (LSB) of pixels in a block are flipped to embed the additional information. Nevertheless, when the legitimate receiver extracts the embedded information or recovers the image, this method may produce errors in some rough regions. To improve this method, Hong $et~al.$ \cite{hong2012improved} proposed a spatial correlation and edge matching mechanism between adjacent blocks to reduce the error rate in image recovery. However, the above two classic RDHEI methods have a common application limitation, that is, the information extraction and the image recovery are inseparable. To solve this problem, Zhang \cite{zhang2011separable} designed a separable RDHEI method. The legitimate receiver can extract the information or recover the image separably based on different keys. Subsequently, various separable algorithms are proposed and further broaden the application scenarios of RDHEI. However, the VRAE method still faces some difficulties and challenges. For example, some VRAE methods may have the bit error rate in information extraction, making it difficult to achieve complete reversibility. Even if the algorithm is reversible, it is prone to a deficiency in limited embedding capacity because of the low redundancy of the encrypted image. The emergence of the RRBE method has improved these situations.

In the second category, the content owner performs processing on the image to reserve room. Then, the image after reserving room is encrypted with the encryption algorithm. Finally, the information hider can embed information in the reserved room of the encrypted image. Ma $et~al.$ \cite{ma2013reversible} put forward an RRBE method for the first time. This method can not only prevent the information about the original image from being leaked but also realize real reversibility. Zhang $et~ al.$ took advantage of prediction error histogram to reserve room before encrypting the image in \cite{zhang2014reversibility}. The additional information can be extracted from the ciphertext or the plaintext according to different needs. As the ciphertext domain without considering the visual quality of the image, Puteaux $et~ al.$ \cite{puteaux2018efficient} proposed an RDHEI method that predicts the most significant bit (MSB) of pixel for the first time. This algorithm greatly improves the embedding capacity of RDHEI. Then, a lossless compression method of bit-planes is proposed in \cite{chen2019high}. This method improves the embedding capacity by utilizing the correlation between pixels of the original image. In \cite{Mohammadi2020High}, the embedding capacity is improved by utilizing the local difference predictor to reserve room in pixel blocks. Recently, Qiu $et~al.$ proposed an RRBE method in \cite{qiu2020Reversible}, which utilizes reversible integer transformation to reserve room for embedding the information. The above-mentioned methods can not only achieve full reversibility but also greatly increase the embedding capacity.

Embedding capacity serves as an important standard to measure the performance of RDHEI and some high-capacity methods \cite{yin2019reversible,wu2019improved,yin2021reversible,Weng2021High,Wang2021Reversible} have been proposed recently.  In \cite{yin2019reversible}, the binary sequence between the original pixel and the prediction pixel of the image is  compared. Then, the processed number of the same bits from MSB to LSB is recorded and marked with Huffman coding to reserve room. In \cite{wu2019improved}, a parameter binary tree is utilized to mark pixels with different prediction errors. In this work, most pixels among the image can be utilized to reserve room so that a high embedding performance can also be obtained. In \cite{yin2021reversible}, an RDHEI method of multiple bit-planes rearrangements is proposed. This method embeds information by dividing and rearranging bit-planes. Then, a hierarchical embedding method is introduced in \cite{Yu2021Reversible}. This method divides prediction errors into different categories and embeds information hierarchically by utilizing prediction errors with different magnitude. Unlike existing methods, the pixel with large magnitude prediction error can also be used to embed information. Thus, a high embedding capacity can be obtained. 

Lately, an RDHEI method based on lossless compression is proposed in \cite{yin2020reversible}. The extended run-length encoding is adopted in this method to compress each bit-plane to reserve room. To protect information of the image, the content owner encrypts the compressed image. Then, the information hider embeds additional information into the encrypted image. Finally, the legitimate receiver extracts the information or recovers the image separably. Experimental results show that the embedding capacity of this method is improved greatly. 

However, the method in \cite{yin2020reversible} utilizes only one encoding for compressing and ignores the distribution characteristics of the bit-plane. Based on it, we put forward an RDHEI method based on bit-plane compression of prediction error. In this work, we adopt a joint encoding algorithm that takes full advantage of the characteristics of the bit-plane to compress bit-planes. Firstly, we calculate the whole prediction error of the image and divide it into blocks of the same size. Then, each bit-plane of prediction error is rearranged and compressed with the joint encoding algorithm to reserve room. Finally, we encrypt the compressed image to protect the image content. For the information hider, the additional information is encrypted and embedded into the reserved room of the encrypted image. With different keys, the legitimate receiver can extract the information or recover the image separably. Compared with state-of-the-art methods, the proposed method possesses real reversibility and greatly improves the embedding capacity.

The proposed method mainly has the following two contributions:
\begin{itemize}
\item {\verb||}A joint encoding algorithm is developed to compress bit-planes in this paper. The joint encoding algorithm utilizes the characteristics of the bit-plane and combines Huffman coding and run-length encoding ingeniously. With this algorithm, the bit-plane can be compressed efficiently and more room can be reserved to embed information.
\item {\verb||}A high-capacity RDHEI method based on bit-plane compression of prediction error is proposed. By compressing the bit-plane of prediction error with the joint encoding algorithm, the proposed method can obtain higher embedding capacity than state-of-the-art RDHEI works.
\end{itemize}

The rest of this paper is organized as follows: Section 2 introduces the proposed joint encoding algorithm in details. Combined with this algorithm, we propose a high-capacity RDHEI method based on bit-plane compression of prediction error in this paper. The detailed process is described in Section 3. Then, in Section 4, plentiful experiments are carried out and corresponding results are analysed to verify the validity of the proposed method. Finally, the conclusion is depicted in Section 5.

\section{The joint encoding algorithm}
The extended run-length encoding is utilized to compress the bit-plane in \cite{yin2020reversible}. This method obtains a good compression effect, but it ignores the distribution characteristics of bit-planes. Therefore, we improve the extended run-length encoding and propose a joint encoding algorithm in this paper. In addition, we also use the bit-plane rearrangement method \cite{chen2019high} to generate the bitstream with more repeat bits, which can further improve the compression effect.

\subsection{Bit-plane rearrangement}

To take advantage of the correlation between adjacent pixels of an image, Chen $et~al.$ proposed a bit-plane rearrangement method in \cite{chen2019high}. In this method, a grayscale image can be represented by eight bit-planes. Firstly, the bit-plane is divided into non-overlapping blocks with the size of ${ t \times t }$. Then, four sorting orders of bit-plane are generated according to different rearrangement modes within and between blocks. The rearrangement type of bit-plane is marked as two bits. The first bit represents the arrangement within the block, $'0'$ represents the arrangement within the block row by row, and $'1'$ represents the arrangement within the block column by column. The second bit represents the arrangement between blocks, $'0'$ and $'1'$ represent the same meanings as above. According to the above operations, four different bitstreams are generated. For example, when $t=2$, the bit-plane is divided into blocks of $2\times2$ and the corresponding rearranged bitstream is shown as Fig.1. Due to the correlation between adjacent pixels, the adjacent bits in the bitstream after rearrangement are often the same, which creates conditions for compressing the bit-plane.
\begin{figure}[ht]
  \centering
  \includegraphics[width=0.28\linewidth]{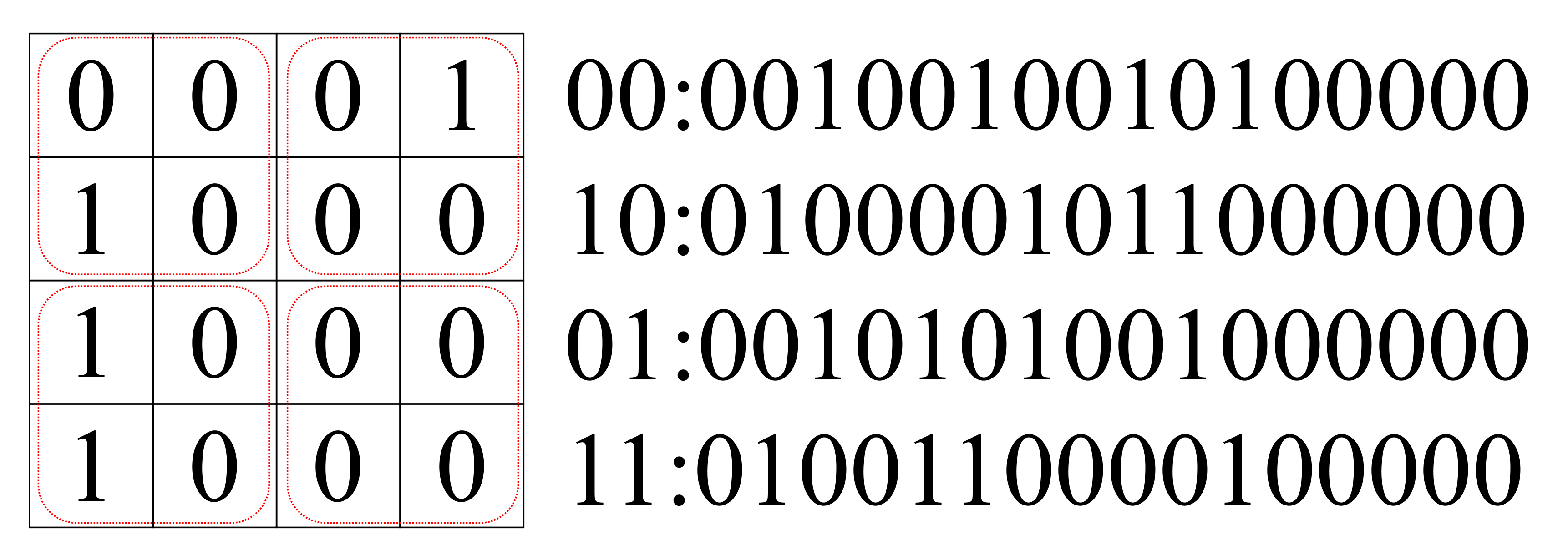}
  \caption{The rearrangement of bit-plane when $t=2$.}
  \Description{The rearrangement of bit-plane when $t=2$.}
\end{figure}

\subsection{Encoding rules}

After the bit-plane is rearranged, the bitstream which contains multiple identical adjacent bits can be obtained. For example, when the bitstream is $'111111111100......011'$, the first repeat bit string $'1111111111'$ consists of ten consecutive $'1'$. We define the length of a repeat bit string as $L$ and compare it with a predefined parameter named $L_{fix}$. If $L\geq L_{fix}$, we treat the current bit string as a long bit string; otherwise, the current bit string is regarded as a short one. The encoding rules are described as follows two cases:
\begin{itemize}
\item {\verb|Case 1|}: When $L<L_{fix}$, the current bit string is judged as a short one and the Huffman coding is utilized. The encoding consists of two parts: the prefix $L_{pre}$ and the Huffman codeword $L_{huff}$. $L_{pre}=1$ means that the current bit string is compressed with Huffman coding. To obtain efficient Huffman coding, short bit strings need to be preprocessed. First, we intercept a bit string whose length is $L_{fix}$ from the first bit of a short bit string. Next, we traverse all bit-planes and record the occurrence probability of each intercepted bit string. Then, the corresponding Huffman coding is generated adaptively with occurrence probabilities to compress the bitstream. Finally, the compressed short bit string can be obtained by connecting $L_{pre}$ and Huffman codeword $L_{huff}$ which corresponds to the intercepted bit string.
\item{\verb|Case 2|}: When $L\geq L_{fix}$, the current bit string is judged as a long one and the run-length encoding is adopted to encode it. The encoding consists of three parts: the prefix $L_{pre}$, the middle part $L_{mid}$, and the suffix $L_{tai}$. $L_{pre}=0$ means that the current bit string is compressed with run-length encoding. $L_{mid}$ represents the binary form of $L$, and the length is determined by a predefined parameter $L_{run}$. For instance, if $L_{run} = 4$ and $L=10$, then $L_{mid}=(1010)_{2}$, where $(*)_2$ represents the binary form of $*$. Finally, the suffix $L_{tai}$ is represented by $'1'$ or $'0'$ to denote the repeat bit.
\end{itemize}

The joint encoding algorithm makes full use of the characteristics of the bit-plane and combines Huffman coding and run-length encoding ingeniously, thus better compression effect can be obtained and more room can be reserved. Take part of the bitstream for example, when $L_{fix}=3$ and $L_{run}=5$, the process of joint encoding algorithm is shown as Fig.2. Firstly, the length of the repeat bit string $L$ is calculated. Next, we compare $L$ with $L_{fix}$ and encode the bit string with the corresponding case. If $L\geq L_{fix}$, the current long bit string is encoded with case 2; otherwise, a bit string with length of $L_{fix}$ is intercepted from the first bit of the current short bit string and compressed with corresponding Huffman coding. Finally, we connect all compressed bit strings to obtain the compressed bitstream. With the above operations, the original bitstream $'00000000000000000101010'$ can be converted to $'01000101111010'$, whose length is far less than the original bitstream. 
\begin{figure}[ht]
  \centering
  \includegraphics[width=0.48\linewidth]{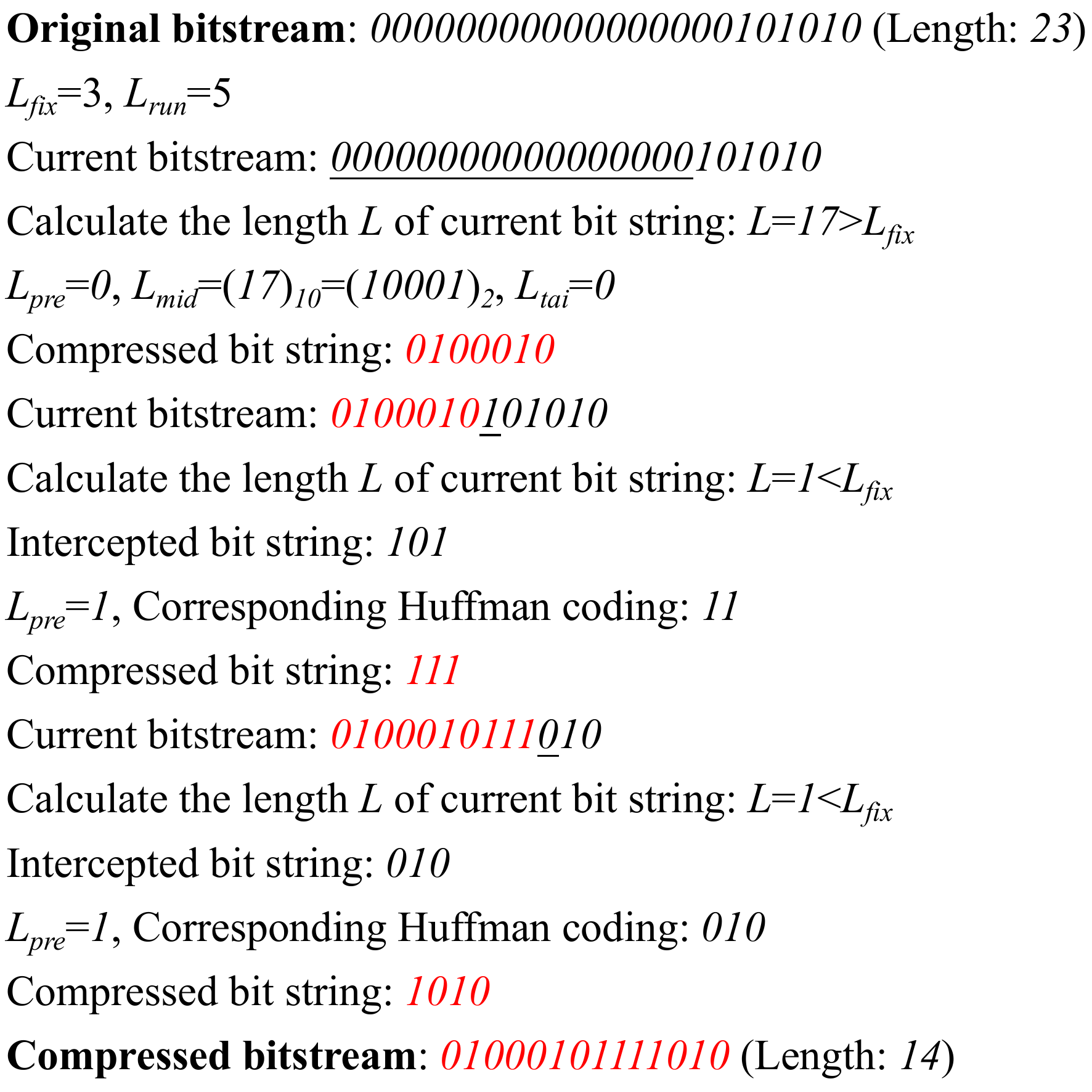}
  \caption{The concrete process of the joint encoding algorithm.}
  \Description{The concrete joint encoding algorithm.}
\end{figure}
\section{RDHEI based on bit-plane compression of prediction error}
To take full advantage of the image redundancy, an RDHEI method based on bit-plane compression of prediction error is proposed in this paper. This section can be described in four parts. Section 3.1 summarizes the overall research framework of the proposed method; In Section 3.2, the content owner reserves room in the original image and encrypts the image; The information is embedded by the information hider in Section 3.3; Finally, the receiver extracts the information or recovers the image based on different keys in Section 3.4.
\subsection{Overall framework}
Usually, the content owner, information hider and receiver are three roles of RDHEI. The content owner masters the original image and encrypts it to protect the image content; The information hider cannot obtain the original image, but he can embed the additional information in the encrypted image; The receiver operates information extraction or image recovery with different keys. Therefore, just as shown in Fig 3, the proposed method can be divided into the following three steps according to different roles:
\begin{itemize}
\item {\verb||}The content owner processes the image to reserve room and encrypts the image. Firstly, the content owner calculates the whole prediction error of the image and divides it into blocks of the same size. Then, each bit-plane of prediction error is rearranged adaptively and compressed with the joint encoding algorithm to reserve room. Finally, the compressed image is encrypted with the image encryption key.
\item{\verb||}The information hider embeds additional information into the encrypted image. The information hider cannot obtain the specific image information from the encrypted image, but he can locate the reserved room. Then, the encrypted additional information can be embedded by bit substitution.
\item{\verb||}The receiver performs information extraction or image recovery with different keys separably. The specific operation is the reverse process of information embedding or image processing.
\end{itemize}
\begin{figure}[ht]
  \centering
  \includegraphics[width=0.75\linewidth]{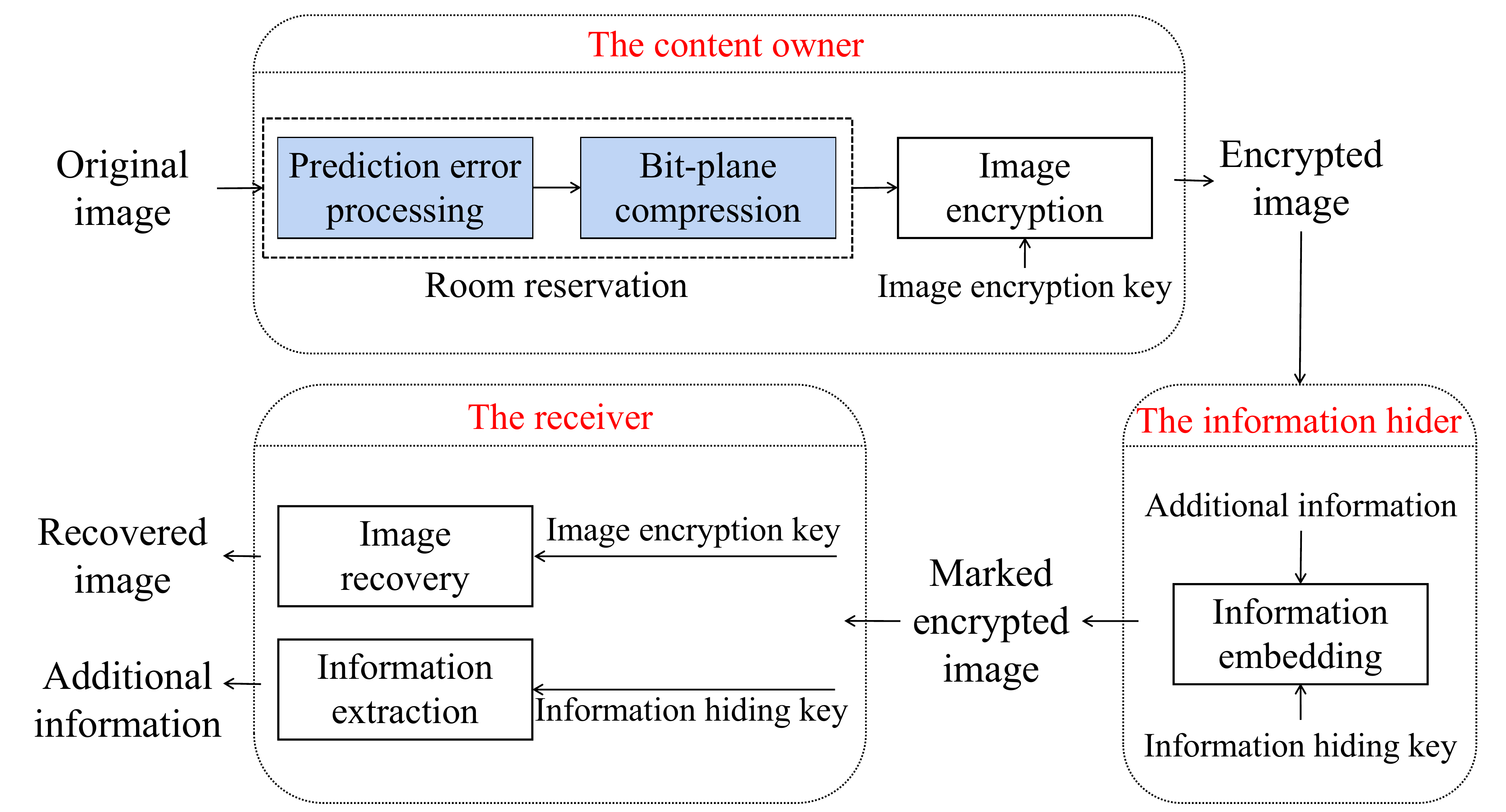}
  \caption{The framework of the RDHEI based on the bit-plane compression of prediction error.}
  \Description{The framework of the proposed method.}
\end{figure}
\subsection{Room reservation and image encryption}
To obtain higher embedding capacity, the content owner reserves room before encrypting the image. Firstly, the content owner calculates the prediction error of the whole image and divides it into blocks. Then, the content owner compresses each bit-plane of prediction error with the joint encoding algorithm to generate the compressed image. After the above operations, the compressed image with the reserved room is obtained. Finally, the content owner encrypts the compressed image to protect the information of the image. Detailed operations are introduced in the remainder of this section.
\subsubsection{Prediction error processing}
\
\newline
For an image $I$ with the size of $m \times n$, the median edge detector (MED) \cite{Weinberger1996LOCO} can be used to calculate the prediction value of the image. $x(i, j)$ represents a random pixel value, where $i,j$ represent the coordinates of the pixel and $2\leq i \leq m$, $2\leq j \leq n$. Three pixels, $x_1$, $x_2$, and $x_3$ that are located on the upper left, the left and the top of $x(i,j)$ are selected to calculate the prediction value $px(i,j)$ with Eq.1. Noticed that the pixels in first row and first column of the image act as the reference pixels without any operations. The remaining pixels are scanned from the second row and second column to calculate the prediction value according to Eq.1.
\begin{equation}
	px(i,j) = \left\{\begin{matrix}
		max(x_{2},x_{3}) & , & x_{1}\leq min(x_{2},x_{3}) \\
		min(x_{2},x_{3}) & , & x_{1}\geq max(x_{2},x_{3}) \\
		x_{2}+x_{3}-x_{1} & , & otherwise
	\end{matrix}\right.
\end{equation}

According to the pixel value $x(i, j)$ and its prediction value $px(i,j)$, the prediction error $e(i,j)$ is calculated as Eq.2. The remaining pixels are scanned sequentially to calculate the whole prediction error of the image.
\begin{equation}
	e(i,j) =x(i,j)-px(i,j)
\end{equation}

Then, the content owner converts the prediction error into the eight-bit binary number with Eq.3, where ${e_k^{'}(i,j)}$ denotes as $k$th bit of the processed prediction error and $\lfloor * \rfloor$ represents the floor function of $*$. Since there are positive and negative prediction errors, the MSB is chosen to represent the sign mark bit. When the prediction error is negative, the MSB is marked with $'1'$; otherwise, it is $'0'$. The lower seven bits are represented by the binary bits of the absolute value of prediction error. Then, the processed prediction error ${e^{'}(i,j)}$ can be calculated by Eq.4. For example, when the prediction error ${e(i,j)=(-100)_{10}}$, the MSB of the processed prediction error should be marked with $'1'$, the lower seven bits are calculated with the absolute of prediction error. With the above descriptions, the prediction error is converted into ${e^{'} (i,j)=(11100100)_2=(228)_{10}}$. Noticed that pixels whose prediction error exceeds [-127,127] are difficult to represent with eight-bit binary number, so these overflow pixels are recorded as auxiliary information. The processed prediction error of overflow pixel is represented by the original pixel value. After the above processing, the distribution of prediction error is more concentrated than the original image so that more room can be reserved.
\begin{equation}
	e_k^{'}(i,j) =  \left\{\begin{matrix}
		            \lfloor \frac{|e(i,j)|}{2^{k-1}}  \rfloor~mod ~2 &,& k=1,2,...,7\\
		            1 &,& e(i,j)<0 &and& k=8\\
		            0 &,& e(i,j)\geq0 &and& k=8\\
		            \end{matrix}\right.
\end{equation}

\begin{equation}
	e^{'}(i,j) = \left\{\begin{matrix}
	             x(i,j)&,&e(i,j)\notin[-127,127]\\
	             {\sum\limits_{k=1}^8 e_k^{'}(i,j)\times 2^{k-1}} &,&others
                 \end{matrix}\right.
\end{equation}

\subsubsection{Bit-plane compression}
\
\newline
A joint encoding algorithm is introduced in Section 2 to compress the bit-plane of prediction error. In the bit-plane compression, Huffman coding can compress short bit strings effectively, and run-length encoding has a better effect on compressing long bit strings. Therefore, the joint encoding algorithm can achieve better compression efficiency than separate encoding. In addition, compared with the original image, the distribution of prediction error performs more concentrated. The compression of prediction error bit-plane can achieve better effect, thus reserving more room. The specific steps of the bit-plane compression are as follows:
\begin{itemize}
    \item {\verb||}The whole prediction error of the image is calculated according to Section 3.2.1. Then, eight bit-planes of prediction error are obtained according to Eq.3 and Eq.4. 
    \item {\verb||}As described in Section 2, each bit-plane of prediction error is rearranged to generate four bitstreams. Then, the content owner compresses bitstreams with the joint encoding algorithm. The bitstream with the best compression effect is adopted and recorded as  ${(P_1,P_2,P_3,P_4,P_5,P_6,P_7,P_8)}$. 
    At the same time, the rearrangement type of each bit-plane is also recorded. The details of bit-plane compression are shown in Algorithm 1.
     \begin{algorithm}
     \SetAlgoLined
     \KwIn{The original bitstream of bit-plane, $L_{fix}$, $L_{run}$}
     \KwOut{The compressed bitstream of bit-plane}
     
     initialization\;
     \While{not at end of this bitstream}{
       calculate the Length $L$ of current bit string with repeat bits\;
       \eIf{$L\geq L_{fix}$}{
         $L_{pre}=0$\;
         $L_{mid}=(L)_2$\;
         $L_{tai}=current~repeat~bit$\;
         }{
         $L_{pre}=1$\;
          intercept bit string whose length is $L_{fix}$ from first bit of current bit string\;
          record each intercepted bit string to generate Huffman coding adaptively\;
          $L_{huff}=current~Huffman~codeword$ 
         }
         connect them ($L_{*}$) to form the current compressed bitstream 
       }
     \caption{Bit-plane compression}
   \end{algorithm}

\begin{figure}[!ht]
	\centering
	\subfigure[]
	{   
		\includegraphics[width=0.5\textwidth]{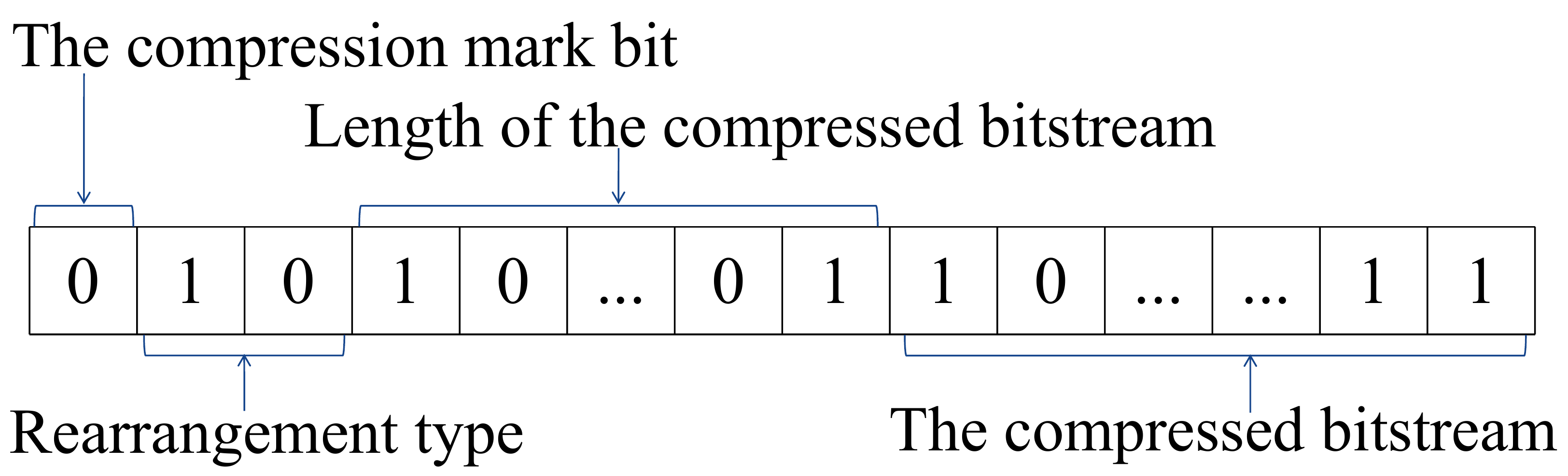}
	}
	\subfigure[]
	{   
		\includegraphics[width=0.5\textwidth]{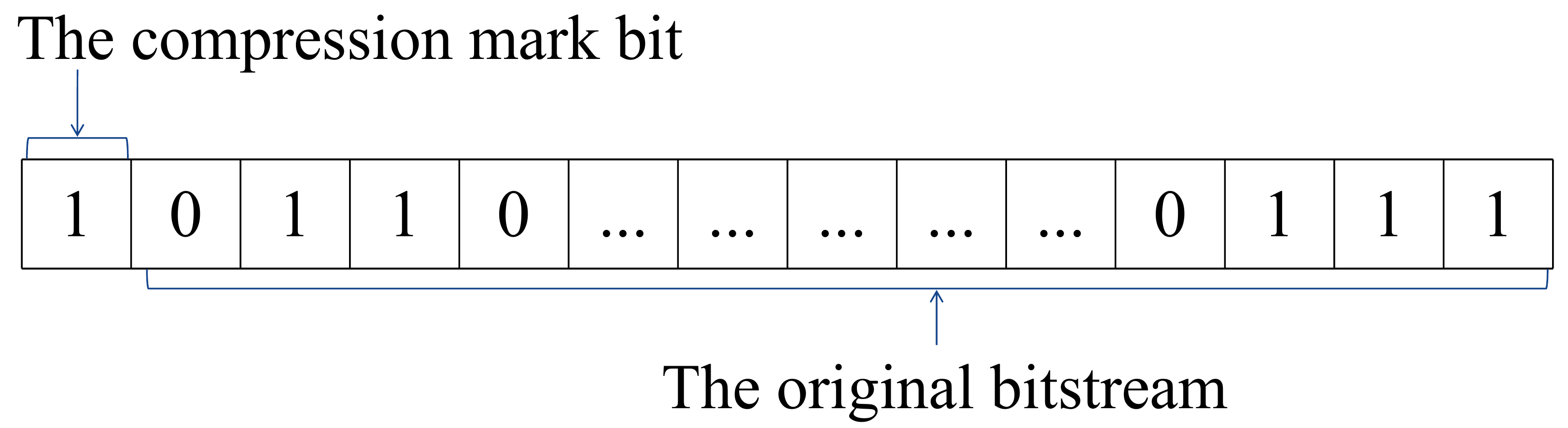}
	}
	\caption{The processed bitstream of the bit-plane: (a) {The compressed information of the bit-plane}; (b) {The uncompressed information of the bit-plane}.}
\end{figure}
    \item {\verb||}After each bit-plane is compressed, the corresponding compressed information can be obtained. Compare the size of the compressed bit-plane and the original bit-plane, if the size of the compressed bit-plane is greater than the original bit-plane, no compression is performed and the uncompressed information is recorded. If less than, the content owner compresses the bit-plane and records the compressed information. As shown in Fig.4(a), the compressed information of the bit-plane consists of the compression mark bit, bit-plane rearrangement type, length of the compressed bitstream, and the compressed bitstream. The compression mark bit is used to judge whether the current bit-plane is compressed or not. When the compression mark bit is $'0'$, it means that the current bit-plane can be compressed; otherwise, the bit-plane cannot be compressed. As shown in Fig.4(b), the uncompressed information of the bit-plane consists of the compression mark bit and the original bit-plane. Finally, all bit-planes are traversed orderly to obtain corresponding compressed or uncompressed information, which is recorded as the processed bitstream of the bit-plane  ${(P_{c1},P_{c2},P_{c3},P_{c4},P_{c5},P_{c6},P_{c7},P_{c8})}$.

    \item {\verb||}Combined with the auxiliary information $A$ which is used for decompressing, all processed bitstreams are connected successively to reconstruct the compressed image ${I_c}$. For the convenience of image recovery, the length of auxiliary information $A$ and the length of the processed bitstreams are recorded in $\log_{2} (m\times n) $ bits and $8 \times \log_{2} (m\times n) $ bits, respectively. As shown in Fig.5, the compressed image ${I_c}$ can be obtained by reconstructing eight bit-planes.
\end{itemize}

For the convenience of decompressing the compressed image, some auxiliary information should be stored in it. The auxiliary information $A$ consists of five parts: The block size $t$, $L_{fix}$, $L_{run}$, Huffman coding rules and information of overflow pixels. 
\begin{figure}[ht]
  \centering
  {
  \includegraphics[width=0.8\linewidth]{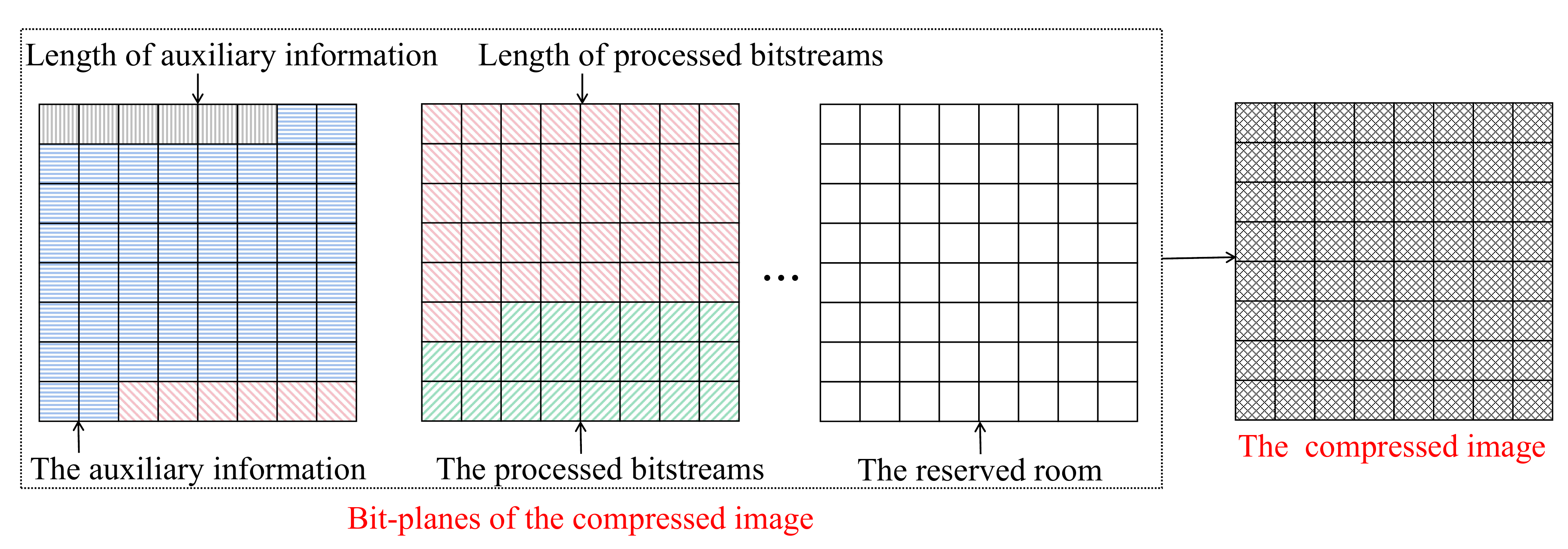}
  }
  \caption{The brief structure of the compressed image.}
  \Description{The brief structure of the compressed image.}
\end{figure}

\subsubsection{Image encryption}
\
\newline
After reserving the room, the content owner encrypts the compressed image ${I_c}$ to prevent the disclosure of image content. In the image encryption stage, firstly, an $m\times n$ pseudo-random matrix $H$ is generated by the image encryption key. Next, both the pixel $x_{c} (i, j)$ of the compressed image ${I_c}$ and the value $h(i, j)$ in the pseudo-random matrix $H$ are converted to their corresponding eight-bit binary form. Formulas for converting are as follows:
\begin{equation}
	x_{c}^{k}(i,j) = \lfloor \frac{x_{c}(i,j)}{2^{k-1}} \rfloor ~mod ~2,k=1,2,...,8
\end{equation}

\begin{equation}
	h^k(i,j) = \lfloor~ \frac{h(i,j)}{2^{k-1}} ~\rfloor ~mod ~2,k=1,2,...,8
\end{equation}

where $x_{c}^{k}(i,j)$ and $h^{k}(i,j)$ represent $k$th bit of $x_{c} (i, j)$ and $h (i, j)$. Then, the content owner implements each bit with exclusive or (XOR) operation to achieve encryption. The encryption formula is as follows:

\begin{equation}
	x_e^{k}(i,j) = x_c^{k}(i,j)\oplus h^{k}(i,j) ,k=1,2,...,8
\end{equation}

\begin{equation}
	x_e(i,j) = \sum\limits_{k=1}^8 x_e^{k}(i,j)\times 2^{k-1}, k=1,2,...,8
\end{equation}

where $x_e^{k}(i,j)$ represents the $k$th bit of the encrypted pixel, $'\oplus'$ represents XOR operation. Then, the encrypted image $I_{e}$ can be obtained by Eq.8, where $x_e(i,j)$ represents the encrypted pixel. Meanwhile, to facilitate the information hider to embed the additional information, the net embedding capacity $c$ is stored in the LSB of the last $8\times \log_{2} (m \times n)$ pixels of the encrypted image by bit substitution.

\subsection{Additional information embedding}
Section 3.2 introduces the specific progress of room reservation and image encryption. With the original bitstream of the bit-plane ${(P_{1},P_{2},P_{3},P_{4},P_{5},P_{6},P_{7},P_{8})}$, the auxiliary information $A$, and the processed bitstream of the bit-plane ${(P_{c1},P_{c2},P_{c3},P_{c4},P_{c5},P_{c6},P_{c7},P_{c8})}$, the net embedding capacity $c$ can be calculated as Eq.9, where $l(*)$ represents the length of $*$.
\begin{equation}
	c =\sum_{k=1}^8 l(P_{k}) - \sum_{k=1}^8 l(P_{ck})-l(A)-17\times \log_{2} (m \times n)
\end{equation}

As described in Section 3.2.3, the net embedding capacity $c$ is stored in the LSB of the encrypted image ${I_e}$. When the information hider receives the encrypted image, he can extract the net embedding capacity and locate the reserved room. Then, the additional information is encrypted with the information hiding key, the encryption is the same as the image encryption in Section 3.2.3. Finally, the information hider embeds the encrypted additional information into the reserved room by bit substitution. In the end, a marked encrypted image $I_{ee}$ is generated. 

\subsection{Information extraction and image recovery}
In the stage of information extraction and image recovery, the receiver extracts the information from the marked encrypted image $I_{ee}$. Firstly, the net embedding capacity is extracted to locate the position of the additional information. Then, the encrypted auxiliary information and the encrypted bitstreams of bit-planes are intercepted from the extracted information. According to different keys held by the receiver, this stage can be divided into three cases:
\begin{itemize}
    \item {\verb||}When the receiver has only the information hiding key, he can only extract the embedded additional information. First, the receiver extracts the net embedding capacity $c$ from the marked encrypted image $I_{ee}$. Then, the position of the additional information is located and the encrypted additional information is extracted. Finally, the receiver decrypts the extracted additional information with the information hiding key. In this way, the additional information can be correctly obtained.
    \item {\verb||}When the receiver has only the image encryption key, he can only recover the image. The receiver extracts the encrypted auxiliary information and the encrypted bitstreams of bit-planes from the marked encrypted image $I_{ee}$. With the image encryption key, the extracted information can be decrypted and the encoding rules can be obtained with the decrypted auxiliary information. Then, the joint encoding algorithm is used to decompress bitstreams to recover the original prediction error. Finally, the original image can be recovered according to Eq.1 and Eq.2.
    \item {\verb||}When the receiver has both the image encryption key and the information hiding key, the embedded additional information can be extracted and the image can be recovered.
\end{itemize}

\section{Experimental results and analyses}
To prove the feasibility, some experiments are carried out in this section. Firstly, we analyze the reversibility and the security in Section 4.1. Then, three parameters are optimized to get better performance. The details are described in Section 4.2. Finally, the performance of the proposed method is compared with state-of-the-art RDHEI methods in Section 4.3. Two well-known grayscale images: Lena and Baboon are used to show the experimental results. Lena is a typical smooth image, while Baboon represents a rough image. To reduce the impact of texture complexity of the image, we also perform the experiment on two databases: BOSSBase \cite{bas2011break} and BOWS-2 \cite{bas2017image}. At the same time, two commonly used indexes, mean square error (MSE) and structural similarity (SSIM) are adopted to test the reversibility of the proposed method. Besides, the embedding rate (ER), which is represented by bpp (bits per pixel) is selected to evaluate the performance of RDHEI. The maximum of the reserved room can be used for obtaining the ER.

\subsection{Analyses of reversibility and security}
Fig.6 shows the image of Lena in different stages. Fig.6(a) is the original image of Lena. Fig.6(b) and Fig.6(c) show the encrypted image and the marked encrypted image, respectively. Fig.6(d) represents the recovered image. It can be seen that the recovered image looks the same as the original image. At the same time, the MSE and SSIM of Fig.6(a) and Fig.6(d) are $'0'$ and $'1'$, respectively, which proves the 
consistency of the recovered image and the original image. That is, the image can be recovered reversibly. To further demonstrate the general effect of this reversibility, we also calculate MSE and SSIM of the recovered image and the original image on two databases: BOSSBase \cite{bas2011break} and BOWS-2 \cite{bas2017image}. In the experimental results, the MSE of images on databases is $'0'$, indicating no difference between recovered images and original images. In addition, the SSIM of images on databases is $'1'$, which represents the same structure of original images and recovered images. That is, the proposed method can recover the image reversibly.

\begin{figure}[!ht]
	\centering
	\begin{minipage}[t]{0.13\textwidth}
		\centering
		\subfigure[]{
			\includegraphics[width=\textwidth]{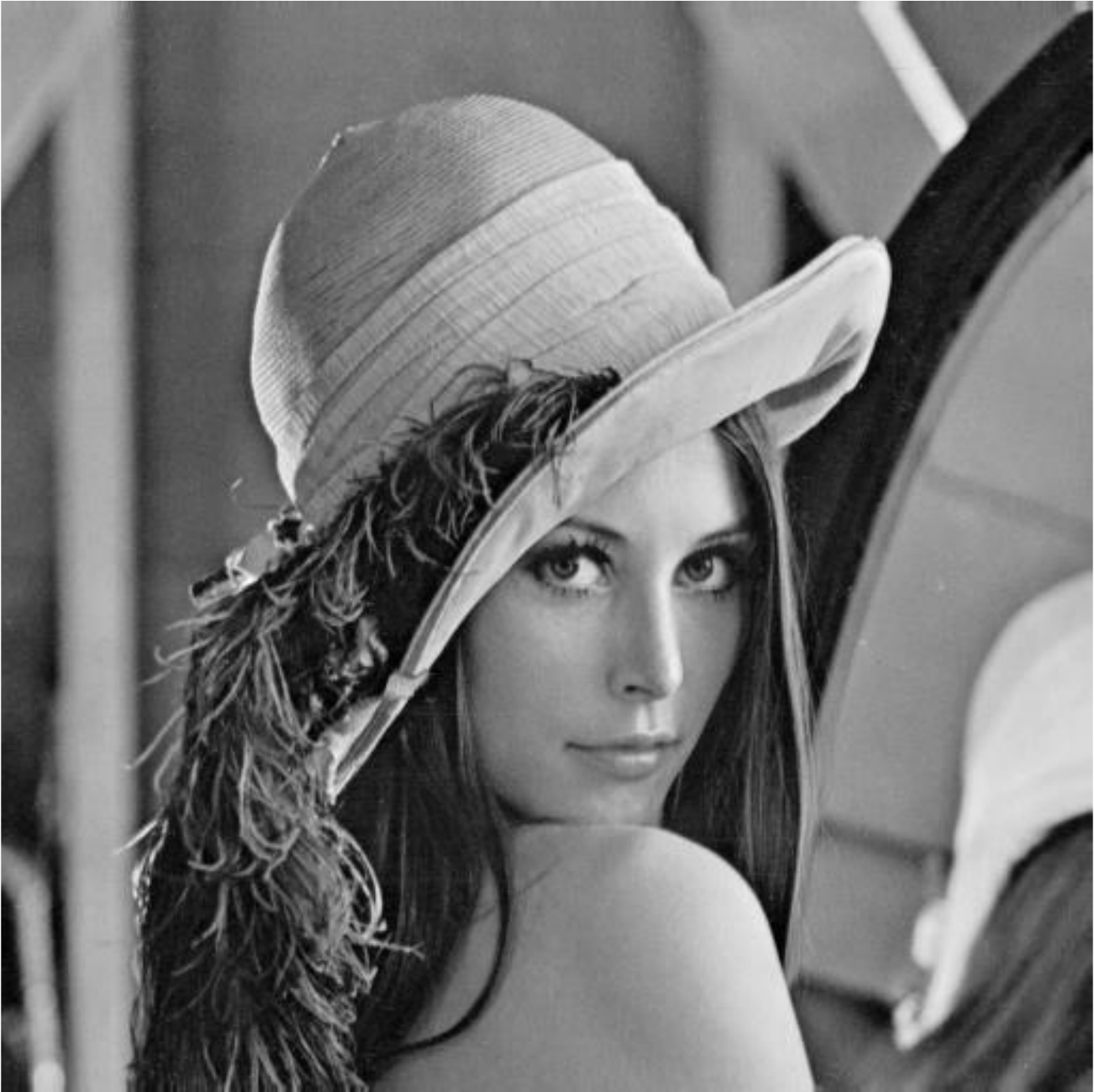}
		}
	\end{minipage}
	\begin{minipage}[t]{0.13\textwidth}
		\centering
		\subfigure[]{
			\includegraphics[width=\textwidth]{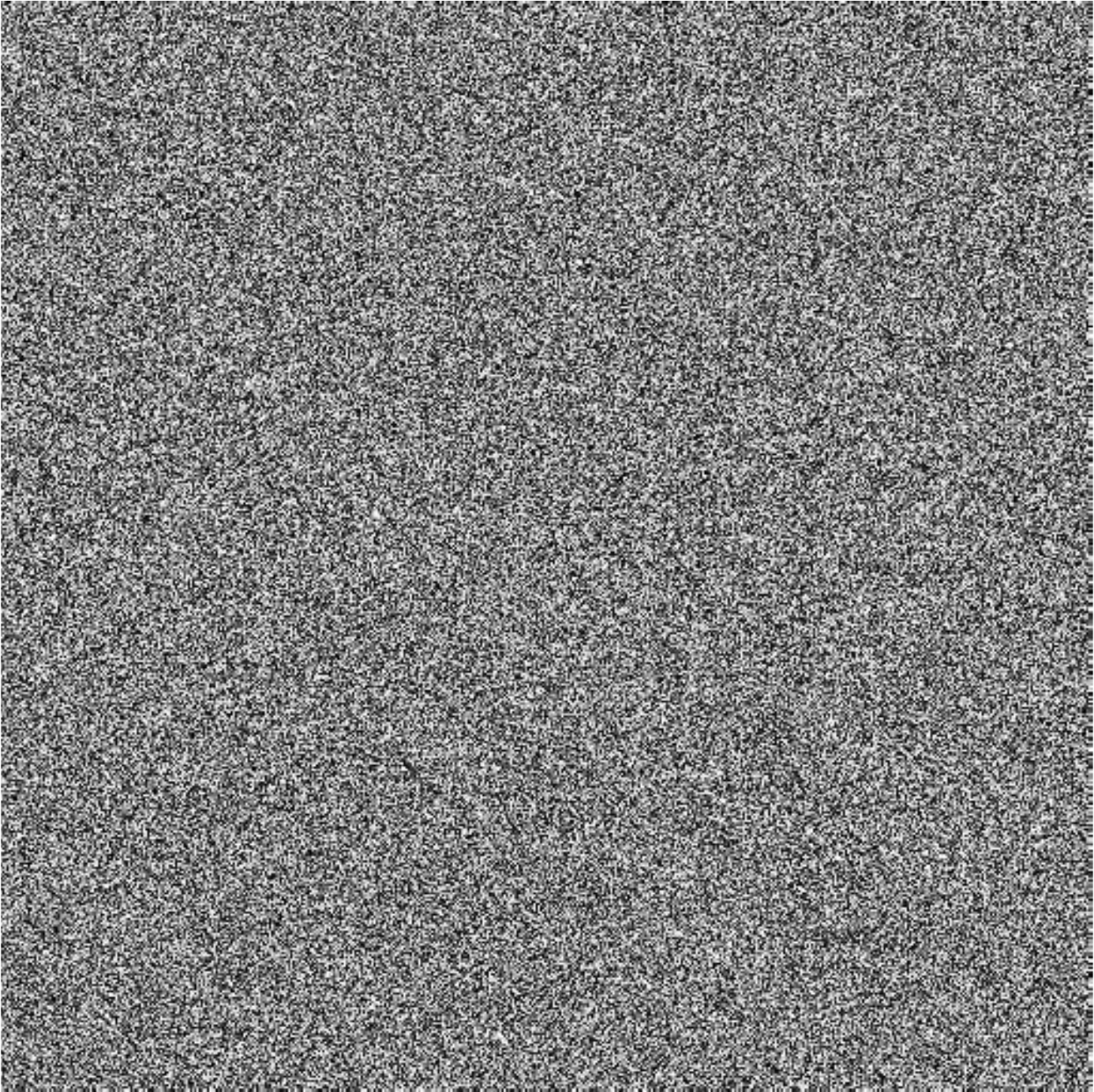}
		}
	\end{minipage}
	\begin{minipage}[t]{0.13\textwidth}
		\centering
		\subfigure[]{
			\includegraphics[width=\textwidth]{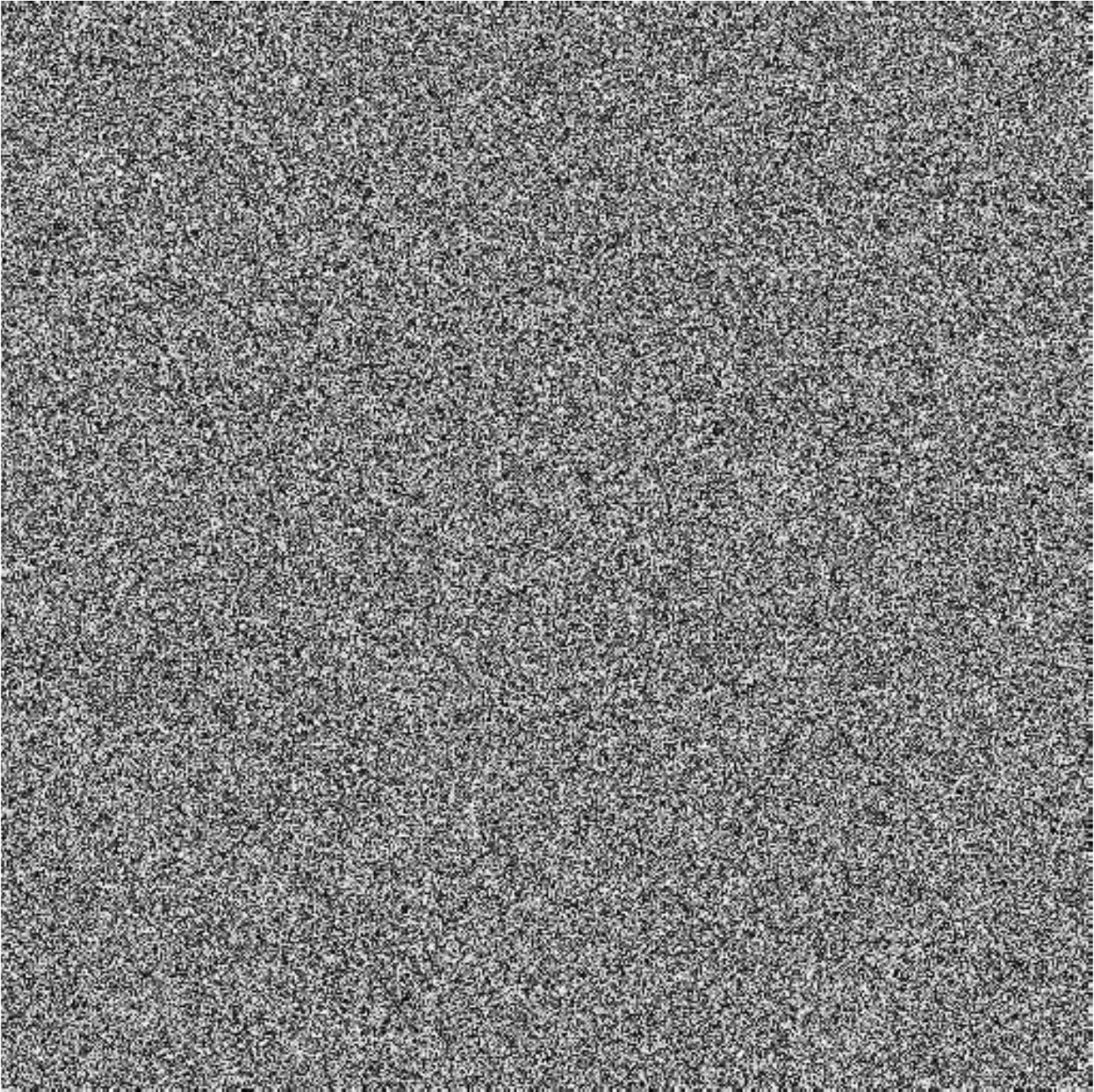}
		}
	\end{minipage}
	\begin{minipage}[t]{0.13\textwidth}
		\centering
		\subfigure[]{
			\includegraphics[width=\textwidth]{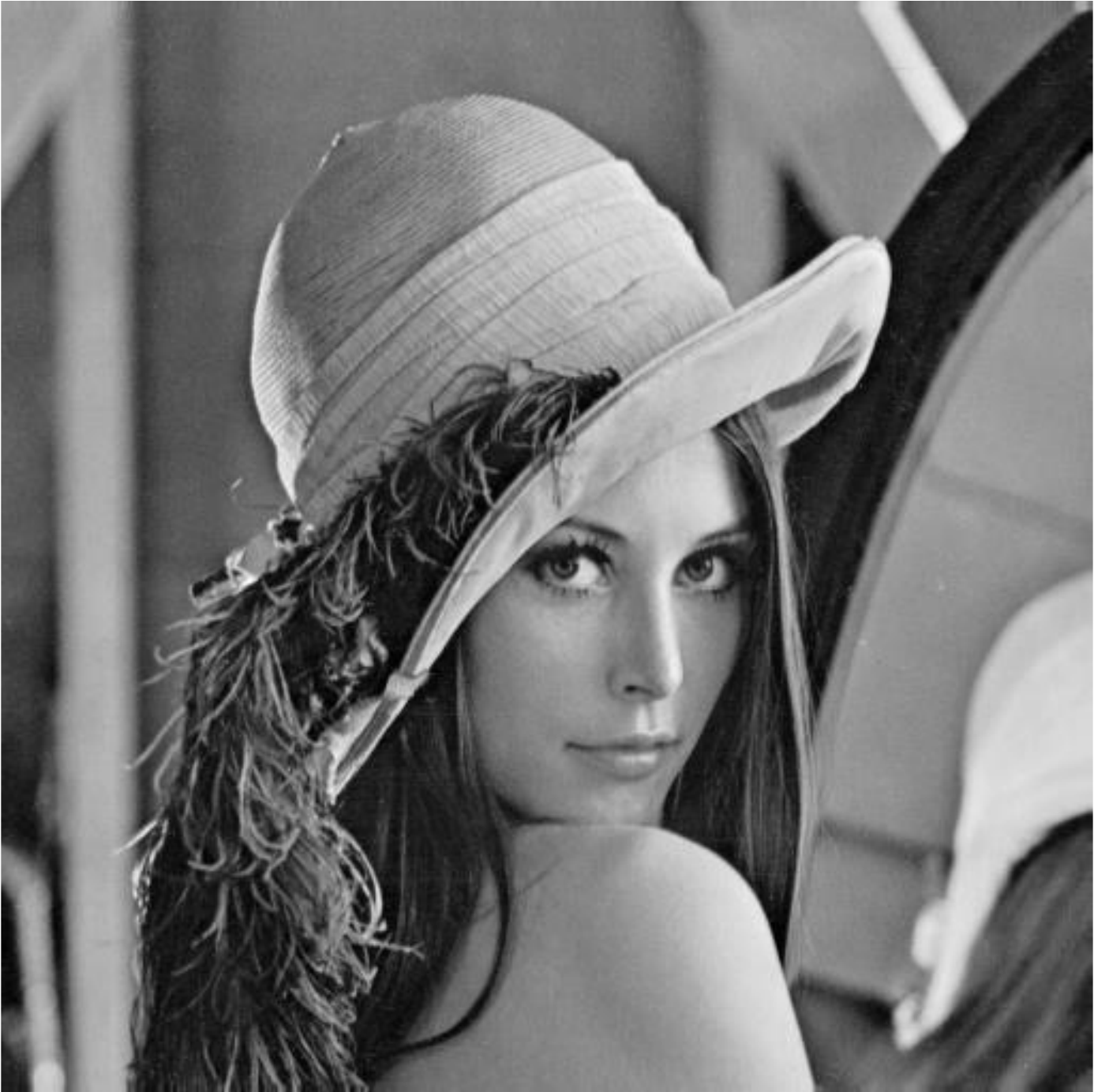}
		}
	\end{minipage}
	\caption{Image of Lena in different stages: (a) The original image; (b) The encrypted image; (c) The marked encrypted image; (d) The recovered image.}
\end{figure}

\begin{figure}[!ht]
	\centering
	\subfigure[]
	{   
		\includegraphics[width=0.38\textwidth]{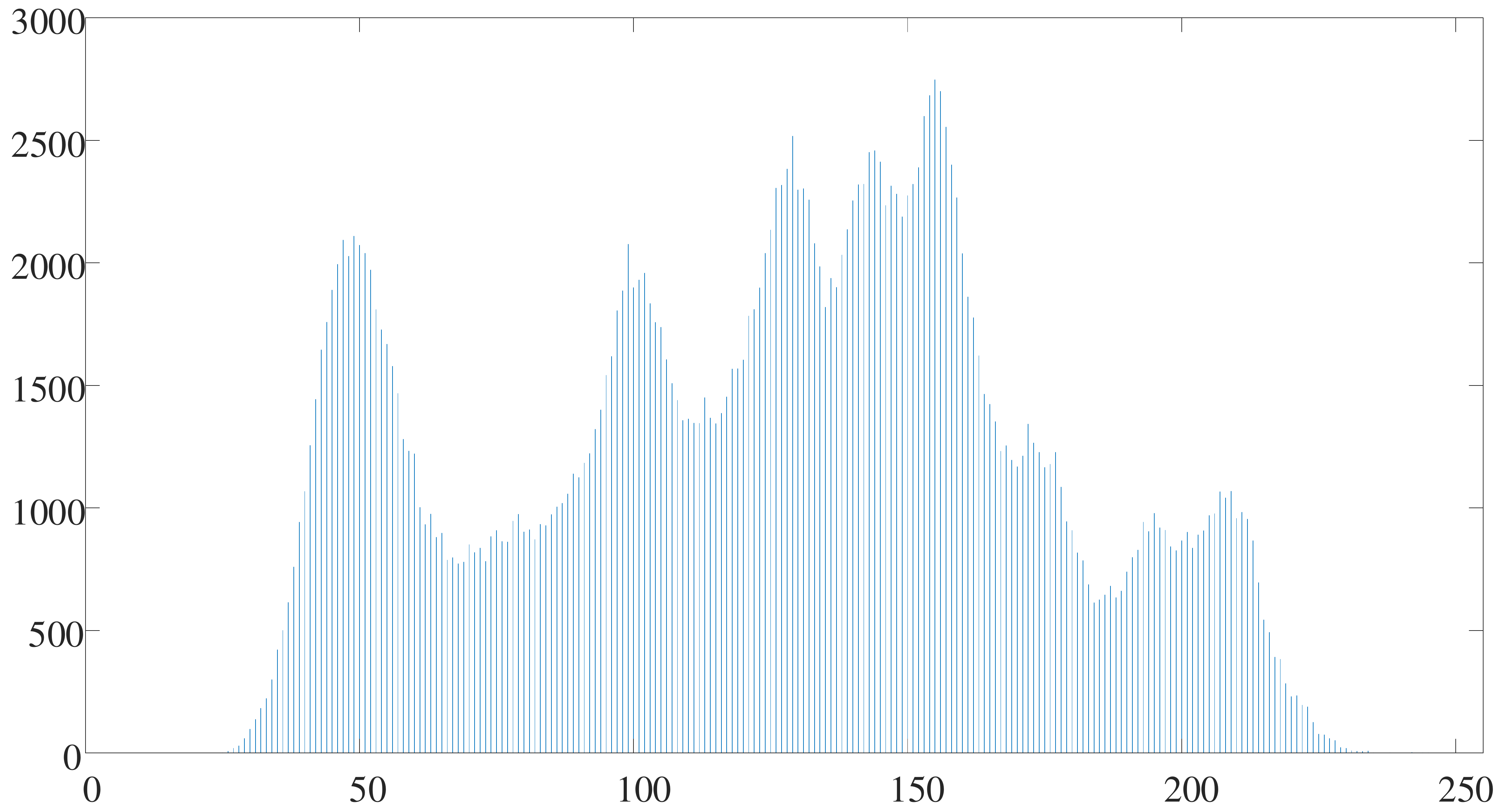}
	}
	\subfigure[]
	{   
		\includegraphics[width=0.38\textwidth]{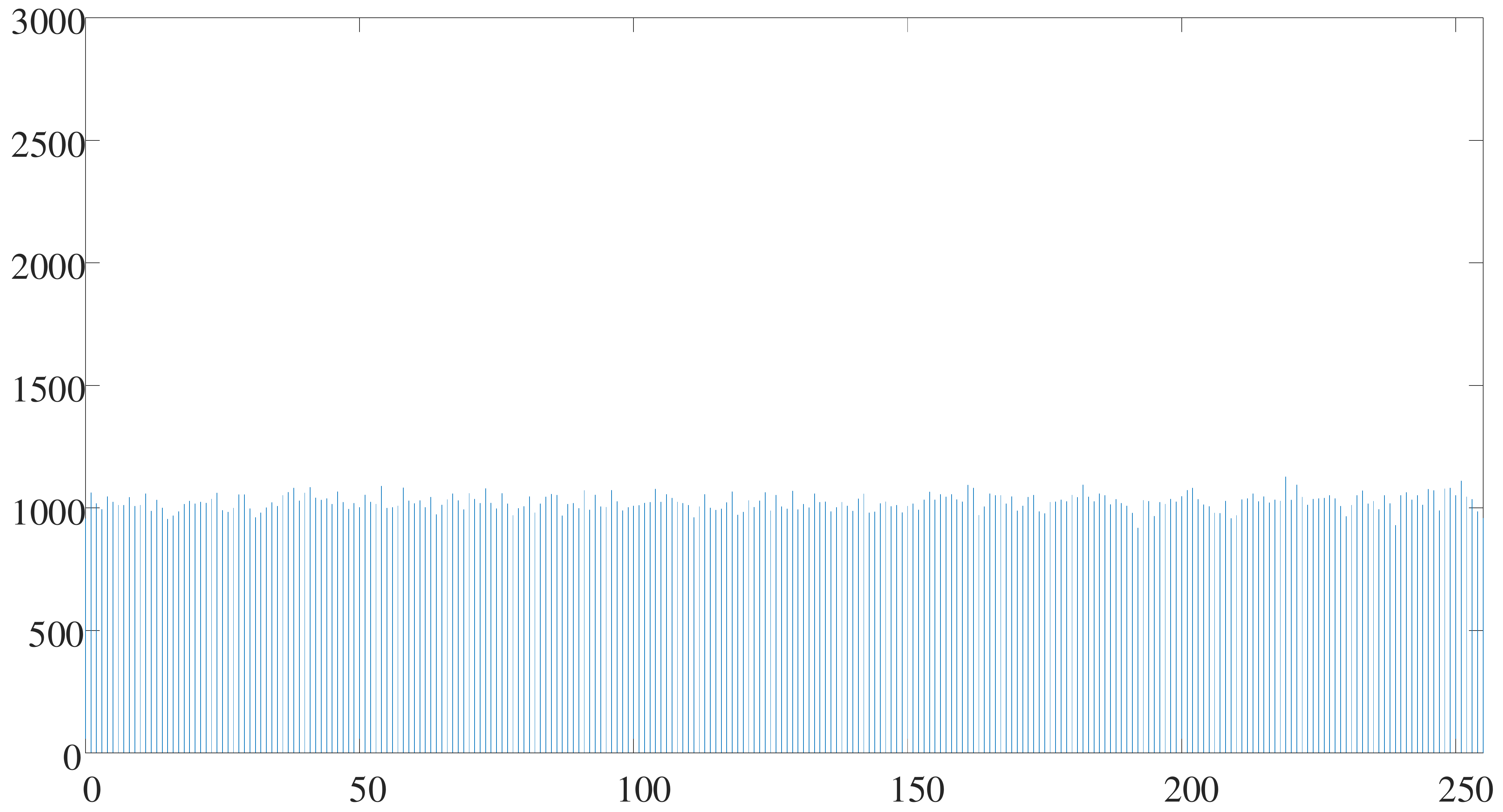}
	}
	\subfigure[]
	{   
		\includegraphics[width=0.38\textwidth]{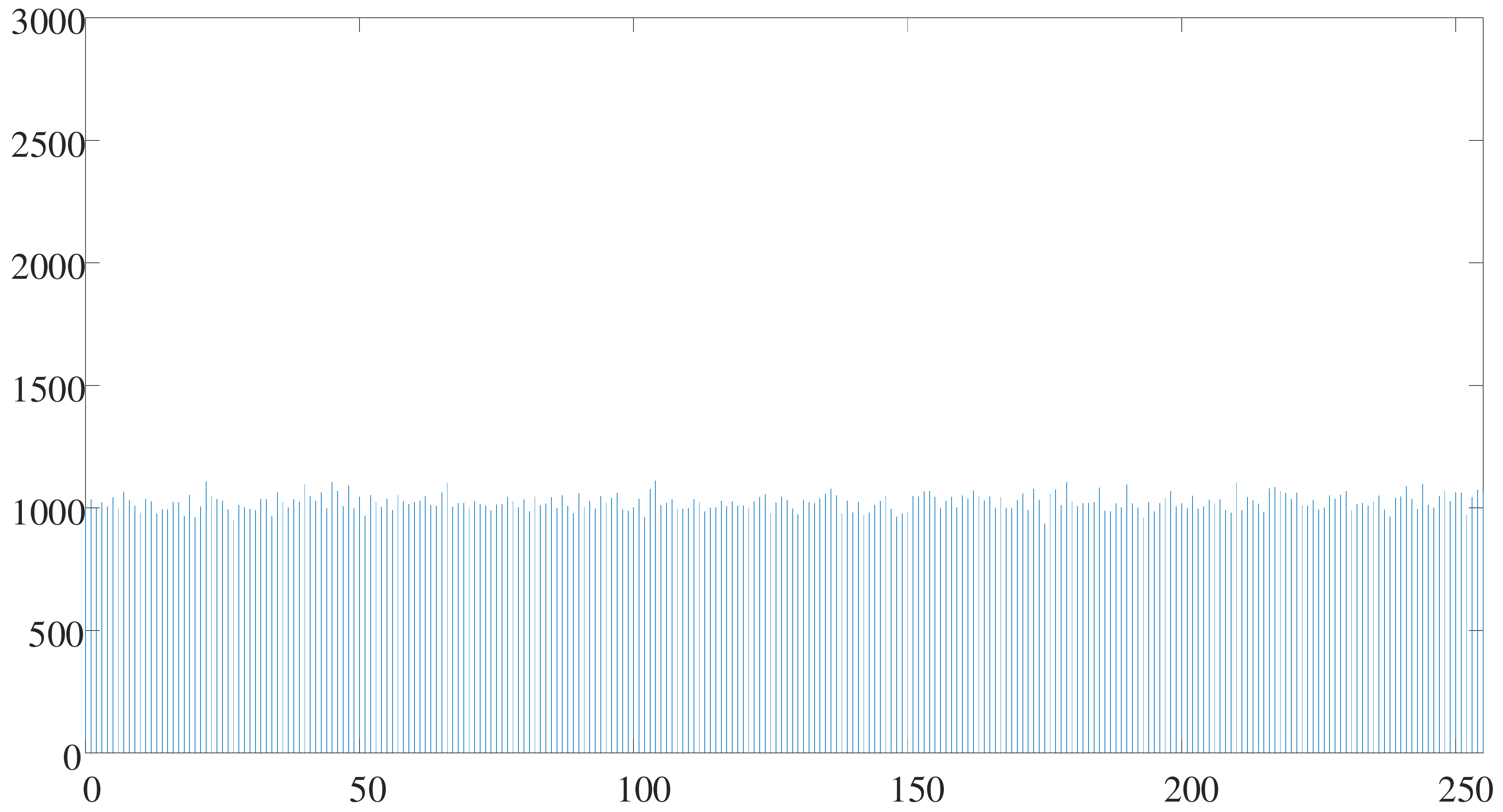}
	}
    \subfigure[]
    {   
    	\includegraphics[width=0.38\textwidth]{Ori-Lena-hist.pdf}
    }
	\caption{ Pixel distribution histograms of Lena in different stages: (a) The original image; (b) The encrypted image; (c) The marked encrypted image; (d) The recovered image.}
\end{figure}

As shown in Fig.6, the encrypted image and the marked encrypted image are meaningless noise images. Thus, it is difficult for an attacker to predict the original image content with them. To verify the security, we analyze the pixel distribution of Lena in different states. Fig.7(a) and Fig.7(d) show the pixel distribution histogram of the original image and the recovered image, respectively. In both images, there are significant pixel fluctuations. Then, the content owner encrypts the image with the image encryption key to protect the image content. Fig.7(b) shows the pixel distribution histogram of the encrypted image, whose pixel distribution is very gentle. Fig.7(c) represents the pixel distribution histogram of the marked encrypted image. Compared with Fig.7(b), Fig.7(c) has no obvious change and the overall pixel distribution is still disorganized. Therefore, the original image of Lena is difficult to obtain based on the encrypted image and the marked encrypted image. In addition, the information hiding key is used to protect the security of the embedded additional information. To sum up, it is difficult for illegal attackers to obtain meaningful information from the ciphertext domain without keys.

\subsection{Parameters optimization}
In the proposed method, there are mainly three parameters: the block size $t \times t$, $L_{fix}$ which is used to judge the type of bit string, and $L_{run}$ which represents the length of run-length encoding. As is described in Section 2.2, a better compression effect can be obtained by making full use of the characteristics of the bit-plane. Thus, more room can be reserved after compression. Obviously, the choice of block size $t \times t$, $L_{fix}$ and $L_{run}$ will affect the compression effect. To optimize parameters, two of three parameters keep unchanged and the rest is modified in the experiment. In the end, parameters with the best performance are selected as the optimal parameters. 
\begin{table}
	\caption{The average ER of 200 images on two databases when $t=4$, $L_{fix}=3$ to $6$, and $L_{run}=3$ to $6$.} 

	\begin{tabular}{cccl}
		\toprule
		$L_{fix}$           & $L_{run}$ & BOSSBase \cite{bas2011break} & BOWS-2 \cite{bas2017image}     \\ 
		\midrule
		\multirow{4}{*}{3}  & 3         & 2.231 & 3.674    \\ 
		
		& 4         & 2.365 & 3.651   \\  
		
		& 5         & 3.385  & 3.593   \\  
		
		& 6         & 3.317 & 3.524   \\ 
		\hline                    
		\multirow{4}{*}{4}  & 3         & 3.516 & 3.718   \\
		
		& 4         & 3.528 & 3.730  \\
		
		& 5         & 3.499 & 3.705  \\
		
		& 6         & 3.454 & 3.662  \\
		\hline
		\multirow{4}{*}{5}  & 3         & 3.533 & 3.735  \\ 
		
		& 4         & 3.562 & 3.764   \\  
		
		& 5         & 3.553 & 3.759   \\  
		
		& 6         & 3.528 & 3.734  \\ 
		\hline                    
		\multirow{4}{*}{6}  & 3         & 3.550 & 3.751  \\
		
		& 4         & 3.589 & 3.789   \\
		
		& 5         & \textbf{3.591} &  \textbf{3.794}   \\
		
		& 6         & 3.576 & 3.780  \\
		\bottomrule
	\end{tabular}
\end{table}

To prevent the texture complexity of the image from affecting the parameter selection, we directly conduct experiments on 200 images from two databases: BOSSBase \cite{bas2011break} and BOWS-2 \cite{bas2017image}. Table 1 describes the average ER of 200 images on two databases when $t=4$, $L_{fix}=3$ to $6$, and $L_{run}=3$ to $6$. It can be seen that when $L_{fix}=6$ and $L_{run}=5$, a higher average ER can be obtained. Beyond this boundary, the average ER shows a decreasing trend. To explore the influence of block size on the average ER, Table 2 describes the average ER of 200 images on two databases when $L_{fix}=6$, $L_{run}=5$, and block size $t \times t$ is $2 \times 2$, $3\times 3$, $4\times 4$, and $8 \times 8$, respectively. It can be seen from Table 2 that when the block size is $4\times 4$, a higher average ER can be achieved. Just as shown in Table 1 and Table 2, better average ER can be obtained when ${t=4}$, $L_{fix}=6$, and $L_{run}=5$. Hence, these parameters are selected as the optimal parameters of the subsequent experiments.

\begin{table}
	\caption{The average ER of 200 images on two databases under different block size when $L_{fix}=6$ , $L_{run}=5$.} 
	\begin{tabular}{ccl}
		\toprule
		Block size & BOSSBase \cite{bas2011break}  & BOWS-2 \cite{bas2017image}  \\ 
		\midrule
		$2 \times 2$    & 3.569 & 3.770 \\  
		$3 \times 3$    & 3.589 & 3.792 \\ 
		$4 \times 4$    & \textbf{3.591} & \textbf{3.794} \\ 
		$8 \times 8$    & 3.566 & 3.772 \\ 
		\bottomrule
	\end{tabular}
\end{table}

\begin{table}
\caption{\label{tb:tab8} The size of compressed bit-plane.} 
\begin{tabular}{ccccc}
\toprule
Bit-plane & Method & Lena & Baboon & Average size\\
		\midrule
\multirow{2}{*}{8$th$} & \cite{yin2020reversible} &/ &/ &/ \\
                       & Proposed  &/ &/ &/ \\
\multirow{2}{*}{7$th$} & \cite{yin2020reversible}&4443 &39985 & 22214\\
                       & Proposed  &3907 &37337 & 20622\\
\multirow{2}{*}{6$th$} & \cite{yin2020reversible}&17933 &143101 & 80517\\
                       & Proposed       &16443 &129710 & 73076\\
\multirow{2}{*}{5$th$} & \cite{yin2020reversible}&58371 &240610 &149490\\
                       & Proposed  &53336 &218774 & 136055\\
\multirow{2}{*}{4$th$} & \cite{yin2020reversible} &161631 &/  & /\\
                       & Proposed &148255 &/ &/\\
\bottomrule
\end{tabular}

\end{table}
\subsection{Performance comparison}
To prove the effectiveness of the proposed method, a large number of comparison experiments are designed in this section. In the comparison experiment, two commonly used standard grayscale images: Lena and Baboon were used. Firstly, the size of each compressed bit-plane is calculated and compared with method \cite{yin2020reversible}. Table 3 shows the size of each compressed bit-plane in two methods, where $'/'$ means that the compressed bit-plane is greater than the original bit-plane and no compression is performed in the current bit-plane. For example, since the $8$th bit-plane represents the sign mark bit of prediction error, it is difficult to obtain a good compression effect due to large fluctuations, lower bit-planes are also the same. From the comparison results, the proposed method can acquire the compressed bit-plane with a smaller size. The main reason is that the joint encoding algorithm takes full advantage of the characteristics of the bit-plane. In the bit-plane compression, the Huffman coding can compress short bit strings effectively, and the run-length encoding has a better effect on compressing long bit strings. Therefore, the joint encoding algorithm can achieve better compression efficiency than separate encoding. By applying this joint encoding algorithm into RDHEI, the proposed method can reserve more room to embed information.

\begin{figure}[!ht]
	\centering
	\subfigure[]
	{   
		\includegraphics[width=0.65\textwidth]{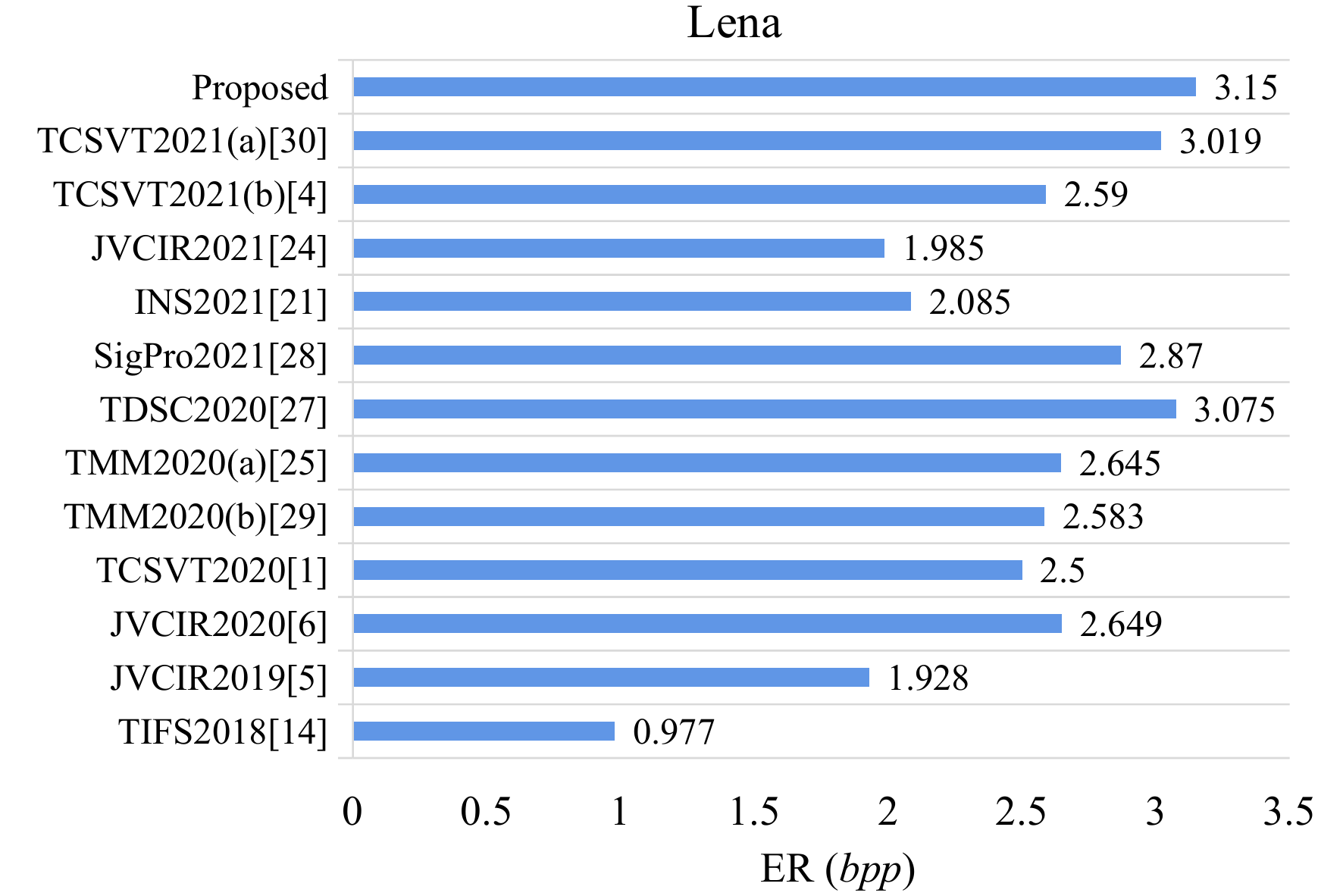}
	}
	\subfigure[]
	{   
		\includegraphics[width=0.65\textwidth]{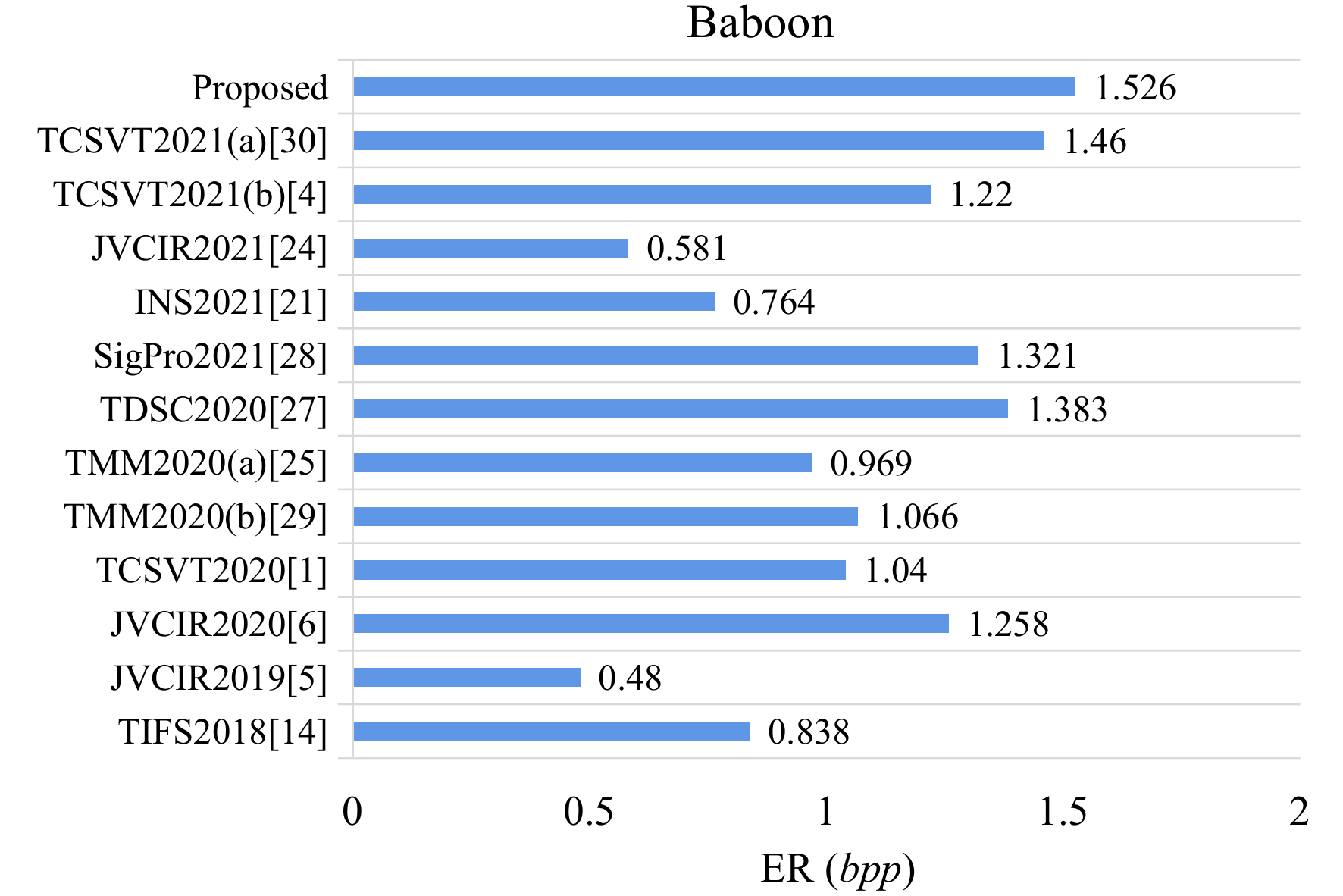}
	}
	\caption{ The ER of test images: (a) {Lena}; (b) {Baboon}.}
\end{figure}

As shown in Section 4.1, the directly decrypted image, that is, the recovered image is consistent with the original image ($MSE=0$ and $SSIM=1$), so this section will not discuss the quality of the directly decrypted image. Then, we compare the ER of the proposed method with state-of-the-art methods \cite{puteaux2018efficient,chen2019high,Guan2020efficient,Mohammadi2020High,yin2019reversible,wu2019improved,yin2020reversible,yin2021reversible,Wang2021Reversible,Weng2021High,Chen2021Multi,Yu2021Reversible} to explore the performance improvement. Unlike other methods, the proposed method adopts a joint encoding algorithm for compression and takes well advantage of the image redundancy to improve the ER. Although some invalid pixels (reference pixels and overflow pixels) cannot participate in reserving room, the effect of such invalid pixels is weeny. Fig.8 shows the comparison results of ER in Lena and Baboon. In Fig.8(a), the ER of \cite{Weng2021High,chen2019high,puteaux2018efficient} is less than 2 $bpp$ while the ER of other methods are improved obviously. In the proposed method, more bit-planes can be compressed adequately to reserve more room. Thus, we can obtain higher ER than state-of-the-art RDHEI methods. Just as shown in Fig.8, the ER of Lena and Baboon reach 3.15 $bpp$ and 1.526 $bpp$, respectively.

\begin{figure}[!ht]
	\centering
    \subfigure[]
    {   
    	\includegraphics[width=0.4\textwidth]{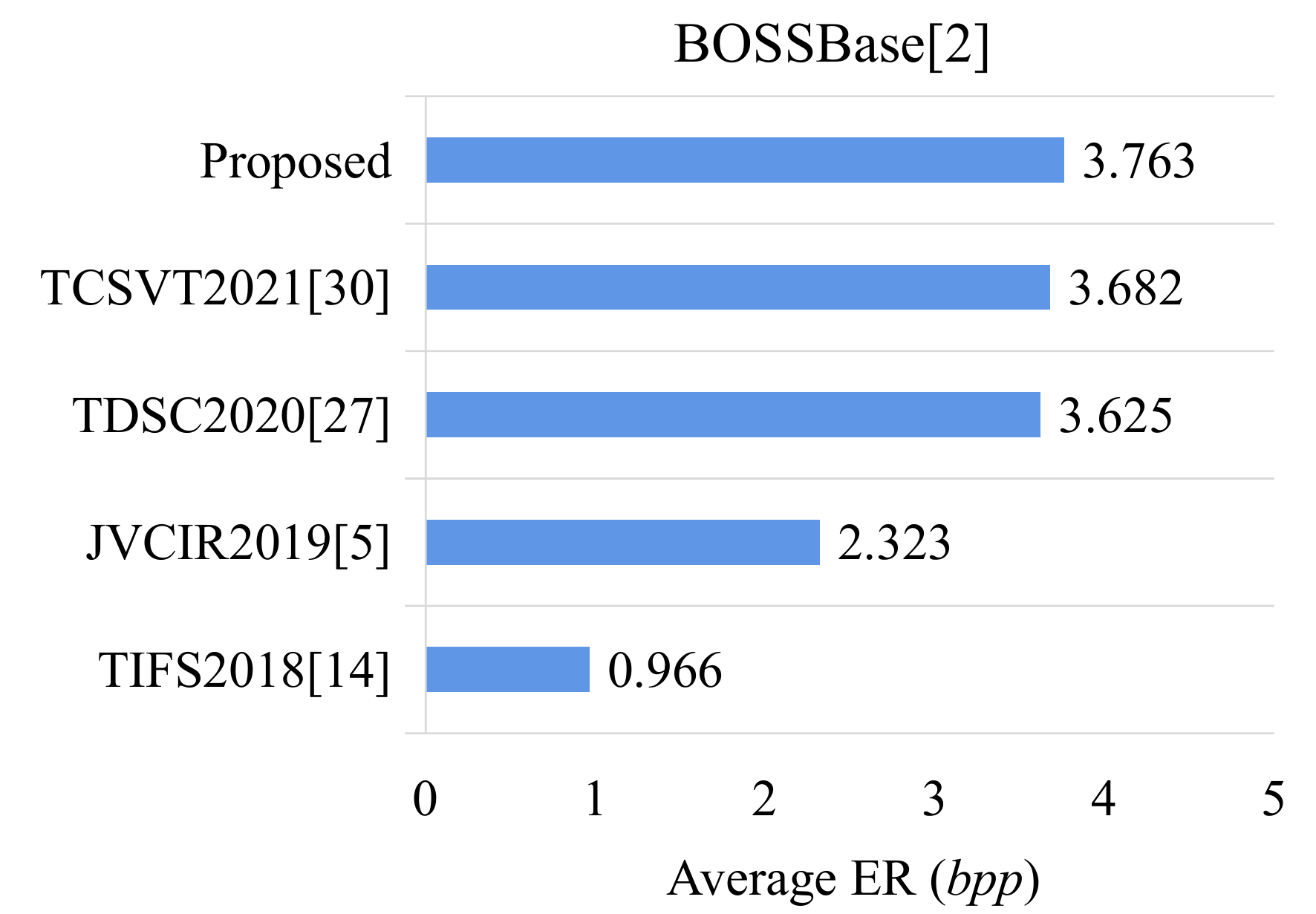}
    }
    \subfigure[]
    {   
    	\includegraphics[width=0.4\textwidth]{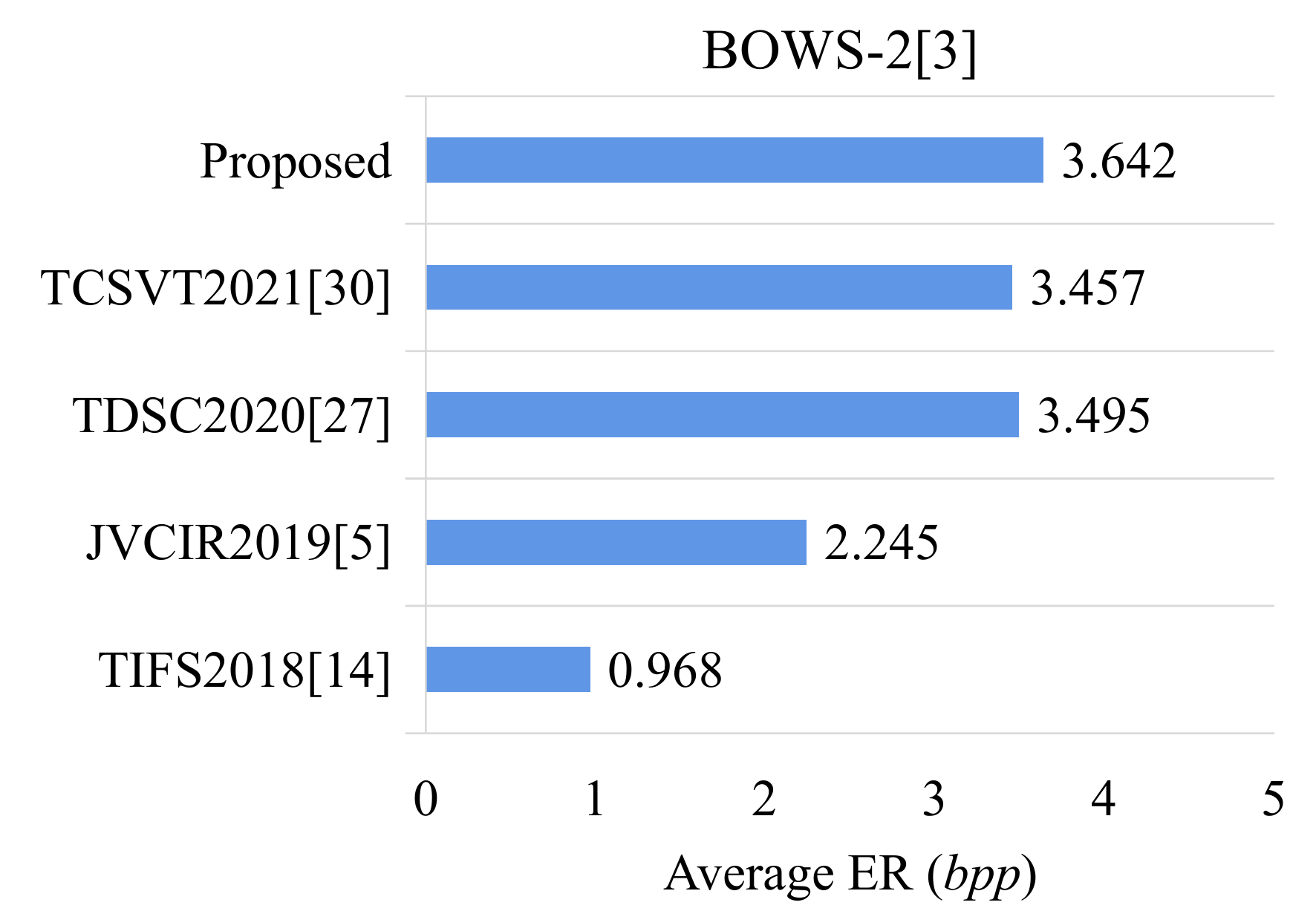}
    }
	\caption{ The average ER on two databases: (a) BOSSBase \cite{bas2011break}; (b)BOWS-2 \cite{bas2017image}.}
\end{figure}

In addition, to reduce the impact of image texture complexity on ER, we also conduct comparison experiments on BOSSBase \cite{bas2011break} and BOWS-2 \cite{bas2017image}. In the comparison experiment on two databases, we select several RDHEI methods \cite{puteaux2018efficient,chen2019high,yin2020reversible,Yu2021Reversible} with the best performance and the most representative. The effectiveness of the proposed method is illustrated by comparing the average ER of all images on databases. Fig.9 shows the comparison results between the proposed method and other methods \cite{puteaux2018efficient,chen2019high,yin2020reversible,Yu2021Reversible}. Similarly, the proposed method can obtain the best average ER on databases. The average ER of the proposed method on BOSSBase \cite{bas2011break} and BOWS-2 \cite{bas2017image} reach 3.763 $bpp$ and 3.642 $bpp$, respectively. Even compared with the highest ER of state-of-the-art methods, the average ER of the proposed method is still improved by 0.081 $bpp$ and 0.147 $bpp$ on two databases, respectively.

\section{Conclusion}
RDHEI plays an important role in information security and privacy protection. To take full use of the image redundancy and improve the embedding capacity of RDHEI, a high-capacity RDHEI method based on bit-plane compression of prediction error is proposed. In this work, we propose a joint encoding algorithm that utilizes the characteristics of the bit-plane. In addition, the bit-plane of prediction error is compressed with the joint encoding algorithm, which further makes well use of the redundancy of the whole image. With these operations, the embedding performance of the proposed method is improved significantly. That is, the embedding capacity of the proposed method is higher than state-of-the-art methods. Moreover, experimental results show that the additional information can be extracted completely and the image can be recovered correctly in the proposed method. 

In the future, we can explore more effective predictors to generate prediction error with concentrate distribution to further improve the embedding capacity. Also, more efficient compression methods can be pursued to improve compression efficiency.

\begin{acks}
This research work is partly supported by National Natural Science Foundation of China (61872003, 61502009).
\end{acks}

\bibliographystyle{ACM-Reference-Format}
\bibliography{sample}


\end{document}